\documentclass{article}
\usepackage{spconf,amsmath,graphicx}
\usepackage{amsfonts}
\usepackage{multirow}
\usepackage{bm}
\usepackage{cite}

\title{Universal ASR: Unifying Streaming and Non-Streaming ASR\\Using a Single Encoder-Decoder Model}
\name{Zhifu Gao$^1$, Shiliang Zhang$^1$,  Ming Lei$^1$, Ian McLoughlin$^2$}
\address{	
	$^1$Speech Lab, Alibaba Group \\
	$^2$ICT Cluster, Singapore Institute of Technology \\
         \{zhifu.gzf, sly.zsl, lm86501\}@alibaba-inc.com, ian.mcloughlin@singaporetech.edu.sg}

\begin{document}
\ninept
\maketitle
\begin{abstract}
Recently, online end-to-end ASR has gained increasing attention. 
\
However, the performance of online systems still lags far behind that of offline systems, with a large gap in quality of recognition. 
For specific scenarios, we can trade-off between performance and latency, and can train multiple systems with different delays to match the performance and latency requirements of various application scenarios. 
\
In this work, in contrast to trading-off between performance and latency, we envisage a single system that can match the needs of different scenarios. 
We propose a novel architecture, termed $Universal~ASR$ that can unify streaming and non-streaming ASR models into one system. 
The embedded streaming ASR model can configure different delays according to requirements to obtain real-time recognition results, while the non-streaming model is able to refresh the final recognition result for better performance. 
\
We have evaluated our approach on the public AISHELL-2 benchmark and an industrial-level 20,000-hour Mandarin speech recognition task. The experimental results show that the $Universal~ASR$ provides an efficient mechanism to integrate streaming and non-streaming models that can recognize speech quickly and accurately. 
On the AISHELL-2 task, $Universal~ASR$ comfortably outperforms other state-of-the-art systems.

\end{abstract}

\section{Introduction}
\label{sec:intro}

Compared to conventional hybrid automatic speech recognition (ASR) systems, 
end-to-end (E2E) systems fold the acoustic, language and pronunciation models into a single sequence-to-sequence model, which dramatically simplifies the training and decoding pipelines. 
\
Currently, there exist three popular E2E approaches, namely connectionist temporal classification (CTC)~\cite{graves2006connectionist}, recurrent neural network transducers (RNN-T)~\cite{graves2012sequence}, and attention based encoder-decoders (AED)~\cite{bahdanau2014neural,chorowski2015attention,chan2016listen}. 
CTC makes an independence assumption that the label outputs are conditionally independent of each other. 
Unlike CTC-based models, RNN-T and AED models have no independence assumption, and can achieve state-of-the-art performance even without an external language model. 

Typical AED models such as LAS~\cite{chan2016listen} and Transformer~\cite{vaswani2017attention}, consist of an encoder and a decoder. 
The encoder transforms raw acoustic features into a higher-level representation,  while the decoder predicts output symbols in an auto-regressive manner. 
The attention module inside the decoder is used to compute soft alignments and produce context vectors. 
\
As originally defined, the soft attention needs to attend to the entire input sequence at each output timestep. 
As a result, soft attention based E2E models are inapplicable to online speech recognition, since they must wait for the entire input sequence to be processed before generating output. 
For real-time ASR services, it is critical to control the latency of the ASR system. 

Recent years have seen much effort spent on building online end-to-end ASR systems. For example, neural transducer (NT)~\cite{sainath2018improving}, Monotonic Chunkwise Attention~(MoChA)~\cite{raffel2017online,chiu2017monotonic,fan2018online,miao2019online}, triggered attention (TA)~\cite{moritz2019triggered,moritz2020streaming}, have all been designed to convert full sequence soft attention into local attention, which is suitable for online speech recognition.\
In~\cite{zhang2020streaming}, Streaming Chunk-Aware Multihead Attention (SCAMA) adopts a predictor to control the encoder output when feeding the decoder, which enables the decoder to generate its output in a streaming fashion. 
\
These efforts have much improved the effectiveness of online E2E ASR systems. 
\
However, the performance of online systems still lags far behind offline systems due to the limited future context information. 
As a general rule, the more future information is available, the better performance can be. 
For real-time applications, we need to carefully trade-off between latency and performance. 
This usually involves training multiple systems with different delays to maximise performance for the given latency requirement of an application scenarios. 
However maintaining multiple versions of ASR systems will obviously significantly increase optimization difficulties as well as operating costs.

Considering the performance gap between online and offline E2E ASR systems, a two-pass E2E system has been proposed in~\cite{sainath2019two,li2020towards,sainath2020streaming}. The two-pass method uses an online RNN-T model to produce streaming predictions, while an offline LAS decoder is used to conduct second-pass rescoring under an acceptable interactive latency. 
Inspired by this, we propose a novel architecture, termed $Universal~ASR$, to unify streaming and non-streaming ASR models into one system. As shown in Fig.\ref{fig:aoa}, both the streaming and non-streaming ASR systems in $Universal~ASR$ are based on the encoder-decoder architecture. The first-pass online ASR system is based on the SCAMA~\cite{zhang2020streaming} model, which contains a latency-control SAN-M~\cite{gao2020san} based encoder and SCAMA based decoder. Moreover, taking the performance and the latency requirements of different application scenarios into account, we propose a Dynamic Latency Training (DLT) strategy that enables the embedded streaming ASR model to configure different delays according to requirements, to obtain real-time recognition results. 
The second-pass non-streaming ASR system based on a SAN-M based encoder-decoder is used to refresh the final recognition result for better performance. Considering the total latency of the system, the first-pass encoder is shared to reduce the overall computational cost of the second-pass encoder. 
Moreover, a stride convolution with down-sampling rate of 2 is introduced to further reduce computational cost. These methods enable the second-pass non-streaming system to generate a final recognition result with acceptable latency. The whole system is jointly trained but can be used either alone or in combination during inference.

We have evaluated our approach on the public AISHELL-2 benchmark (1000-hour) and an industrial-level 20,000-hour Mandarin speech recognition task.
The experimental results show that the $Universal~ASR$ provides an efficient mechanism to integrate streaming and non-streaming models that can recognize speech quickly and accurately. 
On the AISHELL-2 task, $Universal~ASR$ achieved a CER of 5.62\% for the $test\_ios$ test set, which exceeds the state-of-the-art performance for this task.

\section{Methods}
\label{sec:Methods}

\subsection{Overview}
\label{overview}

\begin{figure}[t]
	\centering
	\includegraphics[width=0.85\linewidth]{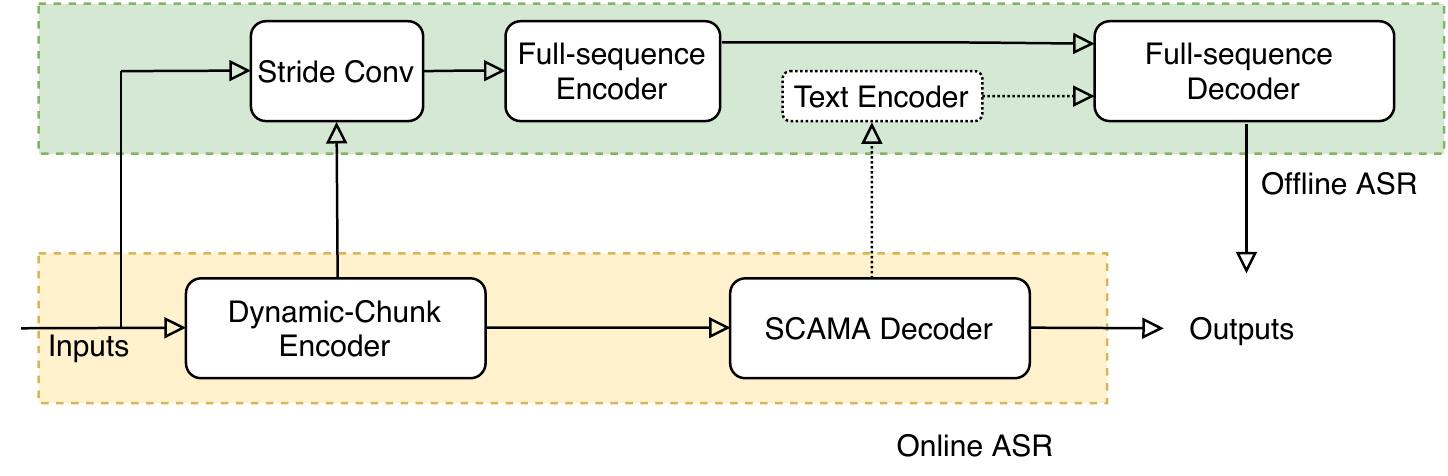}
	\caption{Illustration of $Universal~ASR$ architecture.}
	\label{fig:aoa}
	\vspace{-3mm}
\end{figure}

The proposed $Universal~ASR$ architecture is depicted in Fig.~\ref{fig:aoa}.
The architecture consists of online and offline components, with the two components sharing the dynamic-chunk encoder, to reduce computational complexity.
\
The online components comprise the dynamic-chunk encoder and the SCAMA decoder.
The former adopts an LC-SAN-M structure, whereas the latter is mainly based on a unidirectional deep feed-forward sequential memory network (DFSMN)~\cite{zhang2015feedforward,Zhang2018Deep} with multi-head attention (MHA) layers. The decoder attention mechanism is implemented as SCAMA~\cite{zhang2020streaming}. 
\
The offline components consist of a stride conv, a full-sequence encoder, an optional text encoder and a full-sequence decoder.
The stride conv is a convolution layer with stride of 2.
The full-sequence and text encoders both adopt SAN-M layers, while the full-sequence decoder contains DFSMN and MHA layers.
\ 
The attention mechanism inside full-sequence decoder is composed of one from full-sequence encoder to produce the acoustic context vectors and another from text encoder to produce the semantic context vectors. 
And then the two context vectors are concatenated to produce the attention vectors.

We denote the inputs as $\mathcal{X} = \{\mathbf{X}_1, \mathbf{X}_2,, \cdots, \mathbf{X}_M \}^T$, where $\mathbf{X}_m \in \mathbb{R}^{c \times d}$ are acoustic frames ($d=560$) with chunk-size $c$, and $M$ is the number of chunks in $\mathcal{X}$. 
Each chunk-input $\mathbf{X}_m$ is first passed through a dynamic-chunk encoder and outputs chunk-memory $\mathbf{E}^1_m$. The chunk-memory is then fed into the SCAMA decoder immediately to produce predictions $\mathbf{Y}^1_m$.
After the last chunk-input $\mathbf{X}_M$ is processed by the online component, all the inputs $\mathcal{X}$ and chunk-memory $\mathcal{E}^1$ are first concatenated and then fed into the stride conv.
The outputs of the stride conv are passed through full-sequence encoder to output full-sequence acoustic memory $\mathcal{E}^2$.
Meanwhile, the predictions $\mathcal{Y}^1$ from the SCAMA decoder are fed into the text encoder to output semantic memory $\mathcal{E}^3$. 
At the end, the acoustic memory $\mathcal{E}^2$ and semantic memory $\mathcal{E}^3$ are forwarded to the full-sequence decoder to produce predictions $\mathcal{Y}^2$. $\mathcal{Y}^2$ is used to rectify $\mathcal{Y}^1$, with significant performance improvement.

\subsection{SAN-M}
\label{sec:san-m}

In our previous work~\cite{gao2020san}, we proposed SAN-M to boost the ability of self-attention with a DFSMN memory block.
SAN-M offered an efficient mechanism to incorporate FSMN memory blocks into self-attention to obtain powerful local and long-term dependency modeling abilities.
We will give a brief overview of MHA and SAN-M (detailed in ~\cite{gao2020san}).
MHA~\cite{vaswani2017attention} can be formulated as:
\begin{equation}\label{eq.multihead_self_attention}
{\rm MultiHead}(\mathbf{\mathbf{Q}, \mathbf{K}, \mathbf{V}}) = [{\rm head_{1}, ..., head_{h}}]\mathbf{W}^{O}
\end{equation}
\begin{equation}\label{eq.multihead_self_attention_i}
{\rm head_{i}} = {\rm Attention}(\mathbf{Q}_{i}, \mathbf{K}_{i}, \mathbf{V}_{i}) = {\rm softmax}\left\{\dfrac{\mathbf{Q}_{i}\mathbf{K}_{i}^{T}}{\sqrt{d_{k}}}\right\}\mathbf{V}_{i}
\end{equation}

The FSMN memory block output is calculated as follows,
\
\begin{equation}\label{eq.fsmn_1}
m_t = v_t +\sum\limits_{i = 0}^{L_\ell -1 } a_i \odot v_{t-i} + \sum\limits_{j = 1}^{L_r } c_i \odot v_{t+j} \\
\end{equation}
\begin{equation}\label{eq.fsmn_2}
{\bf M} = [m_1, m_2, \cdots, m_T]^T
\end{equation}
Here, $v_t$ denotes the $t$-th time instance in self-attention values. $L_\ell$ and $L_r$ are the FSMN memory block look-back and look-ahead order respectively. $\odot$ denotes the element-wise multiplication of two equal-sized vectors.

A DFSMN filter has been added on the $values$ inside the MHA to output a memory block. 
The memory content is then added to the output of the MHA, which can be formulated as:
\begin{equation}\label{eq.output}
\mathbf{Y} = {\rm MultiHead}(\mathbf{X}) + {\bf M}
\end{equation}

%

\subsection{LC-SAN-M}
\label{sec:lc-san-m}
In SAN-M, the full sequence attention mechanism renders it unsuitable for streaming. 
In order to control encoder latency, we have extended SAN-M to LC-SAN-M in~\cite{zhang2020streaming}. 
The input sequence $\mathcal{X}$ is divided at chunk-level according to a preset chunk size $c$, denoted as $\mathcal{X} = \{[x_1, \cdots, x_c],[x_{c+1}, \cdots, x_{2c}],\cdots,[x_{nc+1} \cdots x_{T}] \}^T$. The chunk-size $c$ is related to the encoder latency. 
Notationally, $\mathbf{X}_k=\{[x_{kc+1}, \cdots, x_{(k+1)c}]\}^T$ denotes the samples in the $k$-th chunk. 
For each time instance in the $k$-th chunk, it can only access samples in the current and previous chunks. 
Thereby, the output of LC-SAN-M for $\mathbf{X}_k$ can be calculated using the following formulations,
\
\begin{equation}\label{eq.lc_san_m_1}
(\mathbf{Q}_i(k), \mathbf{K}_i(k), \mathbf{V}_i(k)) = (\mathbf{X}_k\mathbf{W}_{i}^{Q}, \mathbf{X}_k\mathbf{W}_{i}^{K}, \mathbf{X}_k\mathbf{W}_{i}^{V})
\end{equation}
\begin{equation}\label{eq.lc_san_m_2}
\bar{\mathbf{K}}_i(k) = [\bar{\mathbf{K}}_i(k-1); \mathbf{K}_i(k)]
\end{equation}
\begin{equation}\label{eq.lc_san_m_3}
\bar{\mathbf{V}}_i(k) = [\bar{\mathbf{V}}_i(k-1); \mathbf{V}_i(k)]
\end{equation}
\begin{equation}\label{eq.multihead_self_attention_i_lc}
{\rm head_{i}}(k) = {\rm SelfAtt}(\mathbf{Q}_i(k), \bar{\mathbf{K}}_i(k), \bar{\mathbf{V}}_i(k))
\end{equation}
\begin{equation}\label{eq.multihead_self_attention_lc}
{\rm MultiHead}(\mathbf{X}_k) =  [{\rm head_{1}(k), ..., head_{h}(k)}]\mathbf{W}^{O}
\end{equation}
\
Furthermore, Eqn. (\ref{eq.fsmn_1}) is modified to the following unidirectional FSMN memory block,
\
\begin{equation}
m_t = v_t +\sum\limits_{i = 0}^{L -1 } a_i \odot v_{t-i}, t\in [kc+1,\cdots,(k+1)c] \\
\end{equation}
\begin{equation}\label{eq.fsmn_2_lc}
{\bf M}_k = [m_{kc+1}, \cdots, m_{(k+1)c}]^T
\end{equation}
Here, $L$ is the total filter order of the FSMN memory block. Finally, we can get the output of LC-SAN-M for $\mathbf{X}_k$ as follows,
\begin{equation}\label{eq.output_lc}
\mathbf{Y}_k = {\rm MultiHead}(\mathbf{X}_k) + {\bf M}_k
\end{equation}

\begin{figure}[t]
	\centering
	\includegraphics[width=0.63\linewidth]{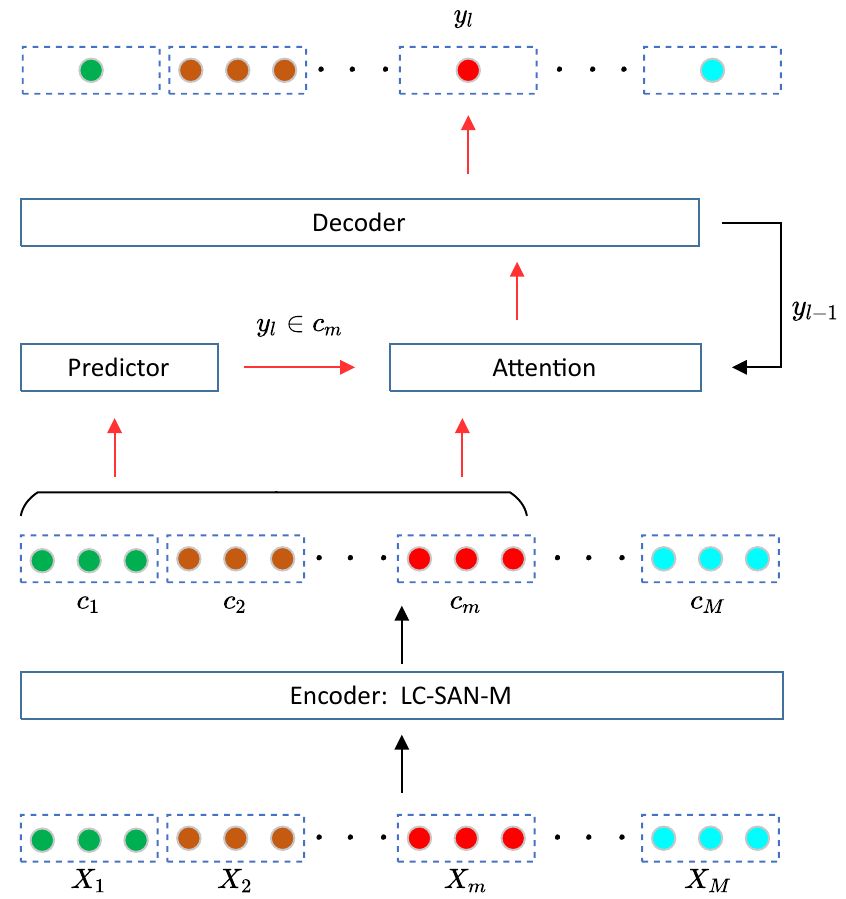}
	\caption{Illustration of SCAMA operation.}
	\label{fig:scama}
	\vspace{-3mm}
\end{figure}

\subsection{SCAMA}
\label{sec:scama}
As illustrated in Fig.~\ref{fig:scama}, we stack a \emph{predictor} on the top of the encoder, which is trained to predict the number of tokens in each chunk. The chunked outputs of the encoder are spliced and then fed into the predictor. Notationally, let us denote the spliced output of the $k$-th chunk as $\mathbf{h}_k^s$. Then the predictor generates the probability $p_k$ as,
\begin{equation}\label{eq.predictor_output}
p_k = {\rm softmax}(\max(\mathbf{h}_k^s\mathbf{W}^1 + \mathbf{b^1},0)\mathbf{W}^2 + \mathbf{b^2})
\end{equation}
The predictor is trained using the cross-entropy loss:
\begin{equation}\label{eq.predictor_ce_loss}
\mathcal{L}_{pred} = -\sum_{k} t_k log (p_k)
\end{equation}
And the overall system is jointly optimized using the following loss function:
\begin{equation}\label{eq.scama_loss}
\mathcal{L} =\mathcal{L}_{e2e} + \alpha \mathcal{L}_{pred}
\end{equation}
where $t_k$ denotes the one-hot vector of the ground truth token number in the $k$-th chunk and $\alpha=0.1$. $\mathcal{L}_{e2e}$ is the original CE-loss to train the encoder-decoder. 
We use a well-trained CTC-based ASR system~\cite{zhang2019investigation} to generate the frame-level alignments and then convert them into the chunk-level labels. 
If the $\ell$-th token of the decoder is in the $m$-th chunk, then only $c_1$ to $c_m$ chunks are fed into the attention module to generate the context vector for the decoder. 
During inference, the class with the maximum probability is chosen as the output for the predictor, which is used to guide how many steps the decoder should attend to the current input chunk.

\subsection{Dynamic Latency Training}
\label{sec:dlt}

Since acceptable latency limits can vary widely for different scenarios, a model with specific latency trained for one task may not satisfy the requirement of other scenarios.
Usually, we need to optimize and maintenance multiple ASR systems, and that would have to be done separately for each. Thus, in this work, we propose a Dynamic Latency Training (DLT) strategy that enables an optimized speech recognition system to be adjusted with different delays according to task requirements. 
For the LC-SAN-M encoder based streaming E2E ASR system, 
the latency is mainly determined by the chunk size. During training, we randomly select a chunk-size from a set, such as $\{5,10,15, 20\}$. During inference, we can flexibly configure the chunk-size according to the latency requirement of a given scenario. In our experiments, we also find that dynamic latency training helps to improve performance, which can be considered a form of regularization.

\begin{table}[t]
	\centering
	
	\caption{Disentangling of Universal ASR (CER\%).}
	\begin{tabular}[t]{c|c|c|c|c}
		\hline
		Structure     & Test Sets & 300ms & 600ms & 900ms \\\hline \hline	
		\multirow{2}*{Universal ASR} 
		& Common & 9.58 & 9.27 & 9.19 \\
		& Far-filed & 14.98 & 14.61 & 14.39 \\\hline
		\multirow{2}*{\quad-\; text-encoder} 
		& Common & 9.76 & 9.53 & 9.46 \\ 
		& Far-filed & 15.03 & 14.78 & 14.56 \\\hline			
		\multirow{2}*{\quad-\; full-encoder} 
		& Common & 10.00 & 9.66 & 9.52 \\ 
		& Far-filed & 15.51 & 15.05 & 14.75 \\\hline	
		
	\end{tabular}
	\label{tab:AOA_offline}
\end{table}

\subsection{System Latency Analysis}
\label{sec:latency}

$Universal~ASR$ offers an efficient mechanism to incorporate the offline model into the online model for various real-time tasks.
For the online model, the latency of the real-time recognition is mainly dependent on the chunk-size of the LC-SAN-M. For example, with chunk-size set to 5, the maximum delay of the online system is about 300ms.
The offline model is used to refresh the final recognition result at an utterance-level. Therefore, it is essential to control the latency introduced by the offline model.
In this work, we adopt two measures to reduce the computational complexity of the offline model in order to generate the final recognition result with acceptable latency. Firstly, since the computational complexity of SAN-M is $\bm{O}(n^{2} \cdot d)$\footnote{$n$ and $d$ are the length and dimension of a sequence respectively.}, we adopt a stride convolution layer with down-sampling rate of $2$ to reduce the computational cost. Secondly, the shared dynamic-chunk encoder enables us to use a smaller full-sequence encoder. In this work, the full-sequence encoder has layer sizes that are half or quarter the size of those in the standard offline E2E model in~\cite{gao2020san}. In our experiments, the real time factor (RTF) of the offline system, with a 20-layer encoder and 10-layer decoder, is about 0.04 on an Intel(R) Xeon(R) CPU E5-2682 v4@2.50GHz.

\begin{table*}[t]
	\centering
	
	\caption{Performance of $Universal~ASR$ on the 20000-hour-task (in CER, \%).}
	\begin{tabular}[t]{c|c|c|c|c|cc|cc|cc}
		\hline
		Model\_id   &	Model0     & Model1  & Model2  & Model3 &  \multicolumn{6}{c}{Model4} \\\hline
		System     & Offline~\cite{zhang2020streaming} & \multicolumn{3}{c|}{SCAMA~\cite{zhang2020streaming}}  &   \multicolumn{6}{c}{Universal ASR}   \\\hline
		\multirow{2}{*}{Latency}   &	\multirow{2}{*}{-}     & \multirow{2}{*}{300ms}  & \multirow{2}{*}{600ms}  & \multirow{2}{*}{900ms} &
		\multicolumn{2}{c|}{300ms}  & \multicolumn{2}{c|}{600ms} & \multicolumn{2}{c}{900ms} \\ \cline{6-11}
		&	     &   &   &  &  online  & offline  & online  & offline & online  & offline \\\hline \hline
		Common & 9.00  & 11.46 & 10.82 & 10.54 & 10.94 & 8.86 & 10.46 & 8.71 & 10.12 & 8.64 \\
		Far-filed & 13.7  & 17.09 & 16.16 & 15.89 & 16.62 & 13.03 & 15.91 & 12.84 & 15.34 & 12.66 \\\hline
		
	\end{tabular}
	\label{tab:AOA_SCAMA}
\end{table*}

\section{Experiments}
\label{sec:exp}

\subsection{Experimental Setup}
We have evaluated the proposed $Universal~ASR$ model on two Mandarin speech recognition tasks, namely the AISHELL-2 task released in~\cite{du2018aishell}, and a 20,000-hour Mandarin task. 
\
For AISHELL-2, we use all the training data (1000 hours) for training, $dev\_ios$ and $test\_ios$ sets for validation and evaluation, respectively. 
\
The 20,000-hour Mandarin task is the same as in~\cite{zhang2019investigation}, which consists of about 20,000 hours of multi-domain data including news, sports, tourism, games, literature, education etc. 
It is divided into training and development sets by a ratio of 95:5. 
A \emph{far-field} set consisting of about 15 hours data and a \emph{common} set consisting of about 30 hours data are used to evaluated the performance.

Acoustic features used in all experiments are 80-dimensional log-mel filter-bank (FBK) energies computed on 25ms windows with 10ms shift.
We stack the consecutive frames within a context window of 7 (3+1+3) to produce the 560-dimensional features and then down-sample the input frame rate to 60ms. 
Acoustic modeling units are Chinese characters, totalling 5211 and 9000 for AISHELL-2 and the 20000-hour tasks respectively. 
As to the detailed experimental setup,  we adopt LazyAdamOptimizer with $\beta_1=0.9$, $\beta_2=0.999$, and the strategy for learning rate is \emph{noam\_decay\_v2} with $d_{model}=512, warmup\_n=8000, k=4$. 
Label smoothing and dropout regularization with a value of $0.1$ are incorporated to prevent over-fitting. 
SpecAugment~\cite{park2019specaugment} is also used in all experiments.

\subsection{20000-hour-task}
\label{sec:20000}

\subsubsection{Performance of Universal ASR}
\label{sec:aoa_on_off}

In this subsection, we evaluate the performance of $Universal~ASR$ on the 20,000-hour-task, detailed in Table~\ref{tab:AOA_SCAMA}.
$Model4$ is our proposed $Universal~ASR$ system.

\
Let us firstly compare the online models.
\
The $Model1$, $Model2$, and $Model3$ are SCAMA online models trained separately with a latency of 300ms, 600ms and 900ms, respectively in~\cite{zhang2020streaming}. 
As expected, the performance becomes better as the latency increases from 300ms to 900ms.
The  online component of $Model4$, which contains a 40-layer encoder and 12-layer decoder, are the same as in~\cite{zhang2020streaming}.
To train $Model4$, we choose $Model1$, as a base-model, and fine tuned it with the DLT enhancement.
From the results, it is clear that DLT can not only enable one model to have different latencies, but also improves its performance.

\
Next we make a comparison with the offline baseline model named  $Model0$. This contains a 40-layer encoder and 12-layer decoder as reported in~\cite{zhang2020streaming}.
The configuration of the offline component of $Model4$ is a stride conv kernel of 5 with stride 2, a 20-layer full-sequence encoder, 10-layer text encoder and 10-layer full-sequence decoder.
Owing to the mechanism of the shared dynamic chunk encoder, the offline component does not need very deep layers compared to $Model0$.
Yet $Model4$, used offline, still outperforms offline $Model0$.

From the results above, we see that the performance of online models still leads offline models by a large gap, due to the lack of future context.
\
The online component of the $Universal~ASR$ model has dynamic latency owing to the DLT, and obtains performance improvement compared to the SCAMA model trained separately.
The offline component has obtained over 15\% and 17\% performance improvement compared to the online component on \emph{common} and \emph{far-field}, respectively.

In summary, besides achieving latency flexibility and outperforming the online models, $Universal~ASR$ also slightly outperforms the offline baseline.

\subsubsection{Disentangling of Universal ASR}

In this subsection, we will explore the contribution of the sub-modules in $Universal~ASR$.
The online component is fine tuned from $Model0$ with DLT and then frozen.
The offline component contains a 10-layer full-sequence encoder, a 8-layer text encoder and a 6-layer full-sequence decoder.
We remove one feature in turn while keeping the others unchanged.
Table~\ref{tab:AOA_offline} reveals the impact of each change to  $Universal~ASR$. 

When we mute the text encoder, the performance decays;
$Universal~ASR$ incorporates the semantic context from predictions of the online component into the offline component via the text encoder to boost performance.
When the full-sequence encoder is removed, the performance obviously worsens by an even greater degree.
The function of the full-sequence encoder is to map the chunk-level acoustic context to a sequence-level context.
Clearly, future context is, as expected, very significant to ASR performance.

\subsection{AISHELL-2}
\label{sec:AISHELL-2}

\begin{table}[t]
	\centering
	
	\caption{Comparison of systems on AISHELL-2 task (CER\%).}
	\begin{tabular}[t]{c|c}
		\hline
		Model                                       & test\_ios \\\hline \hline
		Chain-TDNN~\cite{povey2016purely}           &  8.81   \\
		ESPNET-Transformer~\cite{watanabe2018espnet}&  7.40    \\
		CIF-SAN~\cite{dong2020cif}                  &  5.78  \\\hline
		SAN-M                                       &  5.75  \\
		Universal ASR (online)                      & 6.15 \\
		Universal ASR (offline)                     & 5.62  \\ \hline

	\end{tabular}
	\label{tab:AISHLL2_state_of_art}
\end{table}

The performance of $Universal~ASR$ on AISHELL-2 is detailed in Table~\ref{tab:AISHLL2_state_of_art}.
Chain-TDNN is a popular baseline~\cite{povey2016purely}.
ESPNET-Transformer has been released openly in~\cite{watanabe2018espnet}.
Dong $et.al$ proposed the CIF based Transformer to achieve state-of-the-art performance~\cite{dong2020cif}.
\
We use a SAN-based language model (LM) to perform second-pass rescoring as in~\cite{dong2020cif}.
The hyper-parameter $\alpha$ for LM rescoring is set to $0.3$ while the online component  configurations are the same as in Section~\ref{sec:aoa_on_off}.
The configuration of the offline component is a stride conv kernel of size 5 with stride 2, a 15-layer full-sequence encoder and 10-layer full-sequence decoder.
\
SAN-M is an offline baseline from the present authors~\cite{gao2020san}.
Overall results show that the proposed $Universal~ASR$ model obtains slightly better performance, which is the best reported performance  the authors are aware of to date.

\section{Conclusions}
\label{sec:conclusion}
\
We have proposed a novel architecture, termed $Universal~ASR$ which unifies streaming and non-streaming ASR models into one system. 
The embedded streaming ASR model can configure different delays according to requirements, to obtain real-time recognition results. Meanwhile the non-streaming model is able to refresh the final recognition result for better performance. 

We have evaluated our proposed methods on the public AISHELL-2 benchmark (1000-hour) and an industrial-level 20,000-hour Mandarin speech recognition task.
The experimental results show that $Universal~ASR$ provides an efficient mechanism to integrate streaming and non-streaming models that can recognize speech quickly and accurately. 
When evaluated on AISHELL-2, the proposed $Universal~ASR$ system is able to achieve a CER of 5.62\% on the $test\_ios$ test set, which is state-of-the-art performance for this task.

\vfill\pagebreak

\bibliographystyle{IEEEbib}
\bibliography{strings,refs}

\begin{thebibliography}{10}

\bibitem{graves2006connectionist}
Alex Graves, Santiago Fern{\'a}ndez, Faustino Gomez, and J{\"u}rgen
  Schmidhuber,
\newblock ``Connectionist temporal classification: labelling unsegmented
  sequence data with recurrent neural networks,''
\newblock in {\em Proceedings of the 23rd international conference on Machine
  learning}. ACM, 2006, pp. 369--376.

\bibitem{graves2012sequence}
Alex Graves,
\newblock ``Sequence transduction with recurrent neural networks,''
\newblock {\em arXiv preprint arXiv:1211.3711}, 2012.

\bibitem{bahdanau2014neural}
Dzmitry Bahdanau, Kyunghyun Cho, and Yoshua Bengio,
\newblock ``Neural machine translation by jointly learning to align and
  translate,''
\newblock {\em arXiv preprint arXiv:1409.0473}, 2014.

\bibitem{chorowski2015attention}
Jan~K Chorowski, Dzmitry Bahdanau, Dmitriy Serdyuk, Kyunghyun Cho, and Yoshua
  Bengio,
\newblock ``Attention-based models for speech recognition,''
\newblock in {\em Advances in neural information processing systems}, 2015, pp.
  577--585.

\bibitem{chan2016listen}
William Chan, Navdeep Jaitly, Quoc Le, and Oriol Vinyals,
\newblock ``Listen, attend and spell: A neural network for large vocabulary
  conversational speech recognition,''
\newblock in {\em 2016 IEEE International Conference on Acoustics, Speech and
  Signal Processing (ICASSP)}. IEEE, 2016, pp. 4960--4964.

\bibitem{vaswani2017attention}
Ashish Vaswani, Noam Shazeer, Niki Parmar, Jakob Uszkoreit, Llion Jones,
  Aidan~N Gomez, {\L}ukasz Kaiser, and Illia Polosukhin,
\newblock ``Attention is all you need,''
\newblock in {\em Advances in neural information processing systems}, 2017, pp.
  5998--6008.

\bibitem{sainath2018improving}
Tara~N Sainath, Chung-Cheng Chiu, Rohit Prabhavalkar, Anjuli Kannan, Yonghui
  Wu, Patrick Nguyen, and ZhiJeng Chen,
\newblock ``Improving the performance of online neural transducer models,''
\newblock in {\em 2018 IEEE International Conference on Acoustics, Speech and
  Signal Processing (ICASSP)}. IEEE, 2018, pp. 5864--5868.

\bibitem{raffel2017online}
Colin Raffel, Minh-Thang Luong, Peter~J Liu, Ron~J Weiss, and Douglas Eck,
\newblock ``Online and linear-time attention by enforcing monotonic
  alignments,''
\newblock in {\em Proceedings of the 34th International Conference on Machine
  Learning-Volume 70}. JMLR. org, 2017, pp. 2837--2846.

\bibitem{chiu2017monotonic}
Chung-Cheng Chiu and Colin Raffel,
\newblock ``Monotonic chunkwise attention,''
\newblock {\em arXiv preprint arXiv:1712.05382}, 2017.

\bibitem{fan2018online}
Ruchao Fan, Pan Zhou, Wei Chen, Jia Jia, and Gang Liu,
\newblock ``An online attention-based model for speech recognition,''
\newblock {\em arXiv preprint arXiv:1811.05247}, 2018.

\bibitem{miao2019online}
Haoran Miao, Gaofeng Cheng, Pengyuan Zhang, Ta~Li, and Yonghong Yan,
\newblock ``Online hybrid ctc/attention architecture for end-to-end speech
  recognition,''
\newblock {\em Proc. of Interspeech 2019}, pp. 2623--2627, 2019.

\bibitem{moritz2019triggered}
Niko Moritz, Takaaki Hori, and Jonathan Le~Roux,
\newblock ``Triggered attention for end-to-end speech recognition,''
\newblock in {\em ICASSP 2019-2019 IEEE International Conference on Acoustics,
  Speech and Signal Processing (ICASSP)}. IEEE, 2019, pp. 5666--5670.

\bibitem{moritz2020streaming}
Niko Moritz, Takaaki Hori, and Jonathan Le,
\newblock ``Streaming automatic speech recognition with the transformer
  model,''
\newblock in {\em ICASSP 2020-2020 IEEE International Conference on Acoustics,
  Speech and Signal Processing (ICASSP)}. IEEE, 2020, pp. 6074--6078.

\bibitem{zhang2020streaming}
Shiliang Zhang, Zhifu Gao, Haoneng Luo, Ming Lei, Jie Gao, Zhijie Yan, and Lei
  Xie,
\newblock ``Streaming chunk-aware multihead attention for online end-to-end
  speech recognition,''
\newblock {\em arXiv preprint arXiv:2006.01712}, 2020.

\bibitem{sainath2019two}
Tara~N Sainath, Ruoming Pang, David Rybach, Yanzhang He, Rohit Prabhavalkar,
  Wei Li, Mirk{\'o} Visontai, Qiao Liang, Trevor Strohman, Yonghui Wu, et~al.,
\newblock ``Two-pass end-to-end speech recognition,''
\newblock {\em arXiv preprint arXiv:1908.10992}, 2019.

\bibitem{li2020towards}
Bo~Li, Shuo-yiin Chang, Tara~N Sainath, Ruoming Pang, Yanzhang He, Trevor
  Strohman, and Yonghui Wu,
\newblock ``Towards fast and accurate streaming end-to-end asr,''
\newblock in {\em ICASSP 2020-2020 IEEE International Conference on Acoustics,
  Speech and Signal Processing (ICASSP)}. IEEE, 2020, pp. 6069--6073.

\bibitem{sainath2020streaming}
Tara~N Sainath, Yanzhang He, Bo~Li, Arun Narayanan, Ruoming Pang, Antoine
  Bruguier, Shuo-yiin Chang, Wei Li, Raziel Alvarez, Zhifeng Chen, et~al.,
\newblock ``A streaming on-device end-to-end model surpassing server-side
  conventional model quality and latency,''
\newblock in {\em ICASSP 2020-2020 IEEE International Conference on Acoustics,
  Speech and Signal Processing (ICASSP)}. IEEE, 2020, pp. 6059--6063.

\bibitem{gao2020san}
Zhifu Gao, Shiliang Zhang, Ming Lei, and Ian McLoughlin,
\newblock ``San-m: Memory equipped self-attention for end-to-end speech
  recognition,''
\newblock {\em arXiv preprint arXiv:2006.01713}, 2020.

\bibitem{zhang2015feedforward}
Shiliang Zhang, Cong Liu, Hui Jiang, Si~Wei, Lirong Dai, and Yu~Hu,
\newblock ``Feedforward sequential memory networks: A new structure to learn
  long-term dependency,''
\newblock {\em arXiv preprint arXiv:1512.08301}, 2015.

\bibitem{Zhang2018Deep}
Shiliang Zhang, Lei Ming, Zhijie Yan, and Lirong Dai,
\newblock ``{Deep-FSMN} for large vocabulary continuous speech recognition,''
\newblock pp. 5869--5873, 2018.

\bibitem{zhang2019investigation}
Shiliang Zhang, Ming Lei, Yuan Liu, and Wei Li,
\newblock ``Investigation of modeling units for mandarin speech recognition
  using {DFSMN-CTC-sMBR},''
\newblock in {\em ICASSP 2019-2019 IEEE International Conference on Acoustics,
  Speech and Signal Processing (ICASSP)}. IEEE, 2019, pp. 7085--7089.

\bibitem{du2018aishell}
Jiayu Du, Xingyu Na, Xuechen Liu, and Hui Bu,
\newblock ``Aishell-2: transforming mandarin asr research into industrial
  scale,''
\newblock {\em arXiv preprint arXiv:1808.10583}, 2018.

\bibitem{park2019specaugment}
Daniel~S Park, William Chan, Yu~Zhang, Chung-Cheng Chiu, Barret Zoph, Ekin~D
  Cubuk, and Quoc~V Le,
\newblock ``Specaugment: A simple data augmentation method for automatic speech
  recognition,''
\newblock {\em arXiv preprint arXiv:1904.08779}, 2019.

\bibitem{povey2016purely}
Daniel Povey, Vijayaditya Peddinti, Daniel Galvez, Pegah Ghahremani, Vimal
  Manohar, Xingyu Na, Yiming Wang, and Sanjeev Khudanpur,
\newblock ``Purely sequence-trained neural networks for asr based on
  lattice-free mmi.,''
\newblock in {\em Interspeech}, 2016, pp. 2751--2755.

\bibitem{watanabe2018espnet}
Shinji Watanabe, Takaaki Hori, Shigeki Karita, Tomoki Hayashi, Jiro Nishitoba,
  Yuya Unno, Nelson Enrique~Yalta Soplin, Jahn Heymann, Matthew Wiesner, Nanxin
  Chen, et~al.,
\newblock ``Espnet: End-to-end speech processing toolkit,''
\newblock {\em arXiv preprint arXiv:1804.00015}, 2018.

\bibitem{dong2020cif}
Linhao Dong and Bo~Xu,
\newblock ``Cif: Continuous integrate-and-fire for end-to-end speech
  recognition,''
\newblock in {\em ICASSP 2020-2020 IEEE International Conference on Acoustics,
  Speech and Signal Processing (ICASSP)}. IEEE, 2020, pp. 6079--6083.

\end{thebibliography}

\end{document}